\documentclass[sigconf,nonacm,natbib=true]{acmart}
\AtBeginDocument{%
  \providecommand\BibTeX{{%
    \normalfont B\kern-0.5em{\scshape i\kern-0.25em b}\kern-0.8em\TeX}}}

\copyrightyear{2022} 
\acmYear{2022} 
\setcopyright{acmcopyright}\acmConference[SIGIR '22]{Proceedings of the 45th International ACM SIGIR Conference on Research and Development in Information Retrieval}{July 11--15, 2022}{Madrid, Spain}
\acmBooktitle{Proceedings of the 45th International ACM SIGIR Conference on Research and Development in Information Retrieval (SIGIR '22), July 11--15, 2022, Madrid, Spain}
\acmPrice{15.00}
\acmDOI{10.1145/3477495.3531832}
\acmISBN{978-1-4503-8732-3/22/07}

\usepackage{todonotes}
\usepackage{enumitem}
\usepackage[para]{footmisc}

\begin{document}

\title{On Survivorship Bias in MS MARCO}

\author{Prashansa Gupta}
\affiliation{%
  \institution{University of Glasgow}
  \city{Glasgow}
  \country{United Kingdom}
}
\email{2539575G@student.gla.ac.uk}

\author{Sean MacAvaney}
\orcid{0000-0002-8914-2659}
\affiliation{%
  \institution{University of Glasgow}
  \city{Glasgow}
  \country{United Kingdom}
}
\email{sean.macavaney@glasgow.ac.uk}

\renewcommand{\shortauthors}{Gupta and MacAvaney}

\begin{abstract}
Survivorship bias is the tendency to concentrate on the positive outcomes of a selection process and overlook the results that generate negative outcomes. We observe that this bias could be present in the popular MS MARCO dataset, given that annotators could not find answers to 38--45\% of the queries, leading to these queries being discarded in training and evaluation processes. Although we find that some discarded queries in MS MARCO are ill-defined or otherwise unanswerable, many are valid questions that \textit{could} be answered had the collection been annotated more completely (around two thirds using modern ranking techniques). This survivability problem distorts the MS MARCO collection in several ways. We find that it affects the natural distribution of queries in terms of the type of information needed. When used for evaluation, we find that the bias likely yields a significant distortion of the absolute performance scores observed. Finally, given that MS MARCO is frequently used for model training, we train models based on subsets of MS MARCO that simulates more survivorship bias. We find that models trained in this setting are up to 9.9\% worse when evaluated on versions of the dataset with more complete annotations, and up to 3.5\% worse at zero-shot transfer. Our findings are complementary to other recent suggestions for further annotation of MS MARCO, but with a focus on \textit{discarded} queries. Code and data for reproducing the results of this paper are available in an online appendix.\footnote{https://github.com/seanmacavaney/sigir2022-survivor}
\end{abstract}



\maketitle

\section{Introduction}\label{intro}

Survivorship bias is a type of selection bias where one focuses more on outcomes that pass (survive) a selection process, rather than those that do not~\cite{katopol32maybe}.
Focusing on only the survivors distorts one's view of reality, however. A well-known example of survivorship bias is that
when aircraft return from combat and some are lost, the surviving aircraft should (perhaps unintuitively) be reinforced where damage is \textit{not} apparent, since damage elsewhere is fatal~\cite{abraham-wald}.

Historically, most test collections in information retrieval (e.g., those from TREC) have used a deep pooling technique, where assessors judge many documents for each query (usually hundreds to thousands), pooled from a diverse set of systems~\cite{voorhees2005trec}. This process yields datasets that are relatively complete and can sometimes even accurately measure the performance of ranking algorithms developed decades later (e.g., as seen in~\cite{Voorhees2022CanOT,macavaney-2020-teaching}). However, they are usually limited in the number of queries (usually tens to hundreds), which can reduce the statistical power of experiments that use them~\cite{boytsov-2013-deciding}.
The prominent MS MARCO dataset diverges from this paradigm by assessing far more queries (hundreds of thousands), but with a shallow pooling technique over a single system~\cite{craswell-2021-ms}. We posit that this technique not only biases the ``correct'' answers for each query (a phenomenon studied by~\citet{arabzadeh2021shallow}), but also biases the queries included in the dataset itself; queries that couldn't be found in the top handful of results from a single system are discarded. This is a sizable number of queries, in fact: 38\% and 45\% of queries in the MS MARCO train and dev set, respectively, are discarded. We see this situation as ripe for survivorship bias.

To study the survivorship bias in MS MARCO, we (1) perform an initial qualitative and quantitative analysis of the queries discarded in the MS MARCO passage ranking dataset (Section~\ref{sec:manual}); (2) measure the effect of survivorship bias on the capacity of MS MARCO to provide meaningful evaluations by simulating versions of the dataset that would result in fewer surviving queries (Section~\ref{sec:eval}); and (3) use the simulation approach from (2) to test the effect survivorship has on models \textit{trained} using MS MARCO (Section~\ref{sec:training}).
We find that the majority of the discarded queries in MS MARCO are likely answerable in the available corpus. Further, through extrapolation from our simulation experiments on both training and evaluation, we find that conclusions may differ had MS MARCO contained queries that were discarded. This work motivates further annotation of the MS MARCO dataset with a focus on queries that were previously discarded (e.g., through the TREC Deep Learning Track).

\section{Background, Data, and Scope}\label{survey}

For various parts of our analysis, it is important to understand the annotation process of MS MARCO. Assessors were shown a question-like query and the top 10 passages\footnote{A small minority of queries have more than 10 associated passages, suggesting that assessors were given the option to expand the list if desired.} retrieved by a single system. Based on those answers, the assessor was asked to write a natural-language answer to the question and indicate which passages contributed to it (usually just one passage). If no answer was present in the top results, the question is marked as having ``No Answer Present''. These annotations serve as the basis for a variety of tasks, the most popular of which being the passage and document ranking tasks. In passage and document ranking, the passages (or the documents from which they were extracted) that were marked by assessors are treated as the ground truth; systems are tasked with finding these passages from a corpus of possible answers (8.8M passages or 3.2M documents). Since queries marked as having ``No Answer Present'' have no associated relevant ground truth passages, these queries are discarded for the purposes of passage and document ranking. In other words, queries where the answer could not be found in the top 10 results of a single system (38\% of queries in train and 45\% in dev) are discarded and therefore not used for training or evaluation of ranking systems. Since the document task is simply a derivation of the passage task, we focus our study on the latter.

The passage ranking task data itself does not provide adequate information to thoroughly study this problem. However, we observe that another task derived from the same data -- the MS MARCO QnA task -- provides more complete information. Most importantly, it gives the ordered list of candidate passages shown to the assessors. As we show in Sections~\ref{sec:eval} and~\ref{sec:training}, this information allows us to simulate environments where a more stringent annotation process would be conducted (i.e., one where assessors were shown fewer passages, and therefore fewer queries survive). To further assist in our analysis, we also make use of the automatically-assigned question types, which are also only provided by the QnA task.

We note that there is also the opportunity for survivability bias in other datasets as well. A query (topic) development process is commonplace when constructing IR test collections, such as those from TREC. This process often aims to balance a variety of factors, including assessor interest, difficulty, and estimated number of relevant documents~\cite{voorhees-2004-overview-2,voorhees-2000-overview-1}. 
This process is known to exhibit selection biases, but since the full space of queries considered in these settings are unknown (e.g., they be sourced from private query logs, social media calls, or other processes),  the study of survivorship bias in these settings is challenging~\cite{Melucci2016ImpactOQ}.

\begin{table}
\centering
\caption{Discarded queries by reason they were unanswered.}
\label{tab:inspect}
\scalebox{0.95}{
\begin{tabular}{lrrrrr}
\toprule
Split & \texttt{no\_ans} & \texttt{partial\_ans} & \texttt{wrong\_lab} & \texttt{ill\_form} & Total \\
\midrule
Dev & 70 & 15 & 14 & 1 & 100 \\
Train & 36 & 5 & 6 & 3 & 50 \\
\bottomrule
\end{tabular}
}
\end{table}

Further, a newer version of MS MARCO (named v2) was recently constructed that includes a larger passage corpus~\cite{Craswell2021TrecDl}. This dataset uses an even stricter subset of the MS MARCO passage labels, filtering out those where passages have insufficient textual overlap with any passage in the new corpus. This process resulted in 65\% of queries being discarded (up from 38\% in v1). The survival process here is clearer than the process for v1, since it relates exclusively to changes in online document content over time and changes to the passaging technique used to construct v2. Further, since MS MARCO v2 is not yet widely used for training or evaluation of systems, we focus our study on v1 of the dataset.

Relatedly, we note that a similar survivorship bias exists in the v1 corpus itself: only passages that were shown to assessors are included in the corpus. MS MARCO v2 fixes this problem by including more passages. Therefore, we limit our study to the survivorship bias in terms of queries.

We use \texttt{ir-datasets}~\cite{macavaney-2021-simplified} to access data for our analysis. This package provides access to both the MS MARCO passage ranking dataset\footnote{Dataset ID: \href{http://ir-datasets.com/msmarco-passage}{\texttt{msmarco-passage}}} and the MS MARCO QnA\footnote{Dataset ID: \href{http://ir-datasets.com/msmarco-qna}{\texttt{msmarco-qna}}}. Critically, it also provides a mapping between passages IDs the two datasets (which are not available by default). Matching is conducted by finding textual matches between the two datasets. Unique exact matches are found for all passages once UTF encoding issues are corrected in the passage ranking dataset, so we are confident this process is accurate.

Throughout our analysis, we use four different scoring functions, representative of four different classes of models: sparse lexical (BM25~\cite{robertson-2009-probabilistic}), single-representation (ANCE~\cite{xiong2020approximate}), multi-representation (ColBERT~\cite{khattab-2020-colbert}), and cross-encoder (monoT5~\cite{pradeep2021expandomonoduo}). For neural models (ANCE, ColBERT, monoT5) we use checkpoints released by the authors and model implementations provided by PyTerrier~\cite{macdonald-2021-pyterrier}. To isolate the effect of the model \textit{scoring} function on performance, we use all as re-ranking models over the top 100 BM25 results (using default BM25 parameters).

\section{Analysis of Discarded Queries}\label{sec:manual}

We start our investigation of survivorship bias in MS MARCO by performing a qualitative and quantitative assessment of the discarded queries. We select 100 unanswered queries (i.e., those without any relevance assessments) from the development set and 50 unanswered queries from the training set. We manually inspect these queries along with the top results shown to the assessors (usually 10). Through the inspection, we identify four main situations that occur: (1:\texttt{no\_ans}) the assessor correctly identifies passages do not answer the query, (2:\texttt{partial\_ans}) a passage partially (but not fully) answers the question, (3:\texttt{wrong\_lab}) a passage was presented to the assessor that fully answers the question, and (4:\texttt{ill\_form}) the question is underspecified or otherwise unanswerable.

\begin{table}
\centering
\caption{Number of discarded queries that could be answered by other ranking models.}
\label{tab:inspect2}
\scalebox{0.95}{

\begin{tabular}{lrrrr}
\toprule
 & BM25 & ANCE & ColBERT & monoT5 \\
\midrule
No answer present & 22 (44\%) & 22 (44\%) & 18 (36\%) & 16 (32\%) \\
Answer in top 10  & 28 (56\%) & 28 (56\%) & 32 (64\%) & 34 (68\%) \\
Answer in top 3   & 17 (34\%) & 17 (34\%) & 28 (56\%) & 30 (60\%) \\
\bottomrule
\end{tabular}
}
\end{table}

A summary of the findings from our manual inspection are shown in Table~\ref{tab:inspect}. Importantly, we find that the vast majority of questions themselves are reasonable; only 4 of the 150 queries we inspected are ill-formed. This suggests that there is at least the \textit{potential} for the remaining queries to be answered. We found a relatively low proportion of queries to have been incorrectly labeled as non-answerable (20 of 150, 13\%), which we consider to be reasonable given the subjective nature of the task. The remaining 126 queries (84\%) were correctly labeled as having no answer present (potentially due to incompleteness of the passages given). Overall, these results strongly suggest that the discarded queries from MS MARCO are a result of the shallow pooling technique; most questions are well-formed, the answer simply did not appear in passages shown to the assessor.

Although we now know that most discarded queries are well-formed, we do not yet know whether they could have actually been answered from the available passage corpus. For instance, the questions could be so specific that the answer simply isn't available online, and therefore could not have been part of the web corpus used to construct MS MARCO. To check whether questions reasonably \textit{could} have been answered, we select 50 queries from the prior analysis that were well-formed and not mislabeled. For these queries, we retrieve the top 10 passages from MS MARCO passages using BM25, ANCE, ColBERT, and monoT5. We then label the passages that answer each of the questions from these pools.

Results for this analysis are shown in Table~\ref{tab:inspect2}. We find that all four models are able to identify an answer to a majority of discarded queries within the top 10 results (56--68\%). Note that this serves as a lower bound of the proportion of queries that are potentially answerable in the MS MARCO passage corpus; there could be answers available that were not retrieved by these models. These results suggest that not only are the majority of discarded queries in MS MARCO well-formed, but also answerable in the available corpus.\footnote{We performed the same analysis on the MS MARCO v2 passage corpus, which has over $15\times$ as many passages. We found that even more of the discarded queries were answerable in that corpus (over 75\%).} However, we did observe significant overlap within these pools; when considering all four systems, 15 of the queries remain unanswerable.

Next, we ask whether there are measurable differences between the types of queries that were discarded and those that survived. Table~\ref{tab:type_distribution} breaks down the number of queries by the question type. (Recall from Section~\ref{survey} that these question types are automatically assigned and provided by MS MARCO QnA dataset.) We find that \textsc{Description} and \textsc{numeric} queries are far less likely to survive than \textsc{location} queries. This suggests that \textsc{location} queries are over-represented in MS MARCO, and that, in general, the queries that are discarded come from a different distribution than those that survive. This distribution shift is potentially concerning, given that others have found description queries to be among the more challenging and interesting of the query types~\cite{mackie-2021-deep}.

\begin{table}
\centering
\caption{Distribution of unanswered queries in MS MARCO dev by provided query type label from MS MARCO QnA.}
\label{tab:type_distribution}
\begin{tabular}{lrrr}
\toprule
\textbf{Query Type} & \textbf{Total} & \textbf{Unanswered} & \textbf{(\% of Total)} \\
\midrule
\textsc{description} & 54,616 & 25,519 & (46.7\%) \\
\textsc{numeric}     & 25,547 & 12,048 & (47.2\%) \\
\textsc{entity}      & 8,677  & 3,819  & (44.0\%) \\
\textsc{person}      & 6,425  & 2,487  & (38.7\%) \\
\textsc{location}    & 5,828  & 1,642  & (28.2\%) \\
\bottomrule
\end{tabular}
\end{table}

\section{Survivorship Bias in Evaluation}
\label{sec:eval}

Given that the MS MARCO dev set is commonly used for the evaluation ranking models, we now examine whether the performance of systems may be affected due to survivability bias. To this end, we conduct an experiment where we \textit{simulate} stricter annotation settings based on the data available from the QnA task and extrapolate conclusions from these results. This setting is both low-cost and reproducible (it does not rely on subjective human assessment.)

\subsection{Experimental Setup}
Based on Section~\ref{sec:manual}, we know that a major factor in query survivability is the annotation depth; had annotators inspected more documents, they likely would have been able to answer more of the queries. Collecting a suitable number of annotations for the 45,551 discarded queries in MS MARCO could be potentially very expensive, especially if a greater annotation depth is required for each query. Instead, we simulate settings where lower annotation depths were used and extrapolate from those results to higher depths. We accomplish this simply by filtering the full MS MARCO dev dataset's qrels to only those that occurred in the top $k$ results shown to the annotator. We note that this setup makes the assumption that had a valid answer appeared at a higher rank, the annotator would have selected this answer instead (or selected both) -- a reasonable assumption given well-known rank biases~\cite{joachims-2005-accurately}. We report RR@10 results for BM25, ANCE, ColBERT, and monoT5 using the official MS MARCO evaluation script, and perform statistical testing, with the entire set of qrels as the baseline (representing a null hypothesis that the ranking depth has no effect on the performance of individual queries).

\begin{table}
\centering
\caption{Effect of simulated survivability levels on at different annotation depths in terms of RR@10. Significant differences between the scores of the full MS MARCO dev set (10+) are indicated with * (unpaired t-test, $p<0.05$, Bonferroni corr.)}
\label{tab:eval}
\scalebox{0.9}{
\begin{tabular}{c|rrrr|rr}
\toprule
Depth& BM25  & ANCE  &ColBERT&monoT5 & NumQ & ($\Delta$ NumQ) \\
\midrule
1	&*0.174 &*0.240 &*0.301 &*0.332 &  8,192 & \\
2	&*0.176 &*0.244 &*0.305 &*0.336 & 15,532 & (+7,340) \\
3	&*0.180 &*0.249 &*0.311 &*0.342 & 22,041 & (+6,509) \\
4	& 0.181 &*0.251 &*0.315 &*0.347 & 27,837 & (+5,796) \\
5	& 0.182 & 0.253 &*0.318 &*0.350 & 33,039 & (+5,202) \\
6	& 0.183 & 0.255 &*0.321 &*0.354 & 37,998 & (+4,959) \\
7	& 0.184 & 0.257 & 0.325 &*0.357 & 42,640 & (+4,642) \\
8	& 0.185 & 0.258 & 0.328 & 0.360 & 47,114 & (+4,474) \\
9	&\bf0.187 & 0.259 & 0.330 & 0.363 & 51,309 & (+4,195) \\
10+	&\bf0.187 &\bf0.260 &\bf0.331 &\bf0.365 & 55,542 & (+4,233) \\
\bottomrule
\end{tabular}
}
\end{table}

\subsection{Results}

Table~\ref{tab:eval} presents our experimental results for testing the effect of query survivability on evaluations that use the MS MARCO dev set. The results show that we can reject the null hypothesis that query performance is irrespective of the annotation depth, given that significant differences exist between model performance using only qrels found at lower annotation depths. We observe that this trend is consistent across four classes of models. Perhaps unexpectedly, we find that query performance actually \textit{increases} as deeper judgments are included (e.g., monoT5 performance increases from 0.332 to 0.365, +9.0\% relative improvement).\footnote{One may suspect that this trend could alternatively be explained because more possible relevant documents are introduced to the pool at each step, increasing the chance of a hit. However, we find that this isn't the case, through an experiment where we kept only the first qrel per query and observed the same trend.} While this may be expected for models that employ neural networks (which could be overfit to particular annotation artefacts), we find that the trend also exists for the unsupervised BM25 model, though to a lesser extent (0.174 to 0.187, +7.5\% relative improvement). Extrapolating these results to further depths we can conclude that the reported model performance is likely to differ had the collection been annotated to a further depth. This suggests MS MARCO may be a poor estimate of the \textit{actual} performance of a ranker over a natural selection of queries. However, we note that all models we examine exhibit roughly the same behaviour and the \textit{relative} ranking of systems remains the same. This mirrors findings that suggest similar behaviour is present with more qrels per query (both by means of the TREC DL datasets~\cite{craswell-2019-overview,craswell-2021-ms} and by automatically conducting annotation using relevance feedback~\cite{Mackenzie2021ASA}). However, our study is limited by comparing only four systems that each have a relatively large absolute difference performance; an experiment that makes use of more systems (as was done in the aforementioned studies) is needed to provide conclusive evidence that the ranking of systems is stable.

\section{Survivorship Bias in Training}\label{sec:training}

Given our observation that survivability has a noticeable effect on the \textit{evaluation} of systems that use MS MARCO, it is natural to ask whether systems \textit{trained} on it are also affected by its survivability bias. This is of particular concern given that it has become commonplace to perform either zero-shot transfer from MS MARCO to other datasets (e.g.,~\cite{pradeep2021expandomonoduo,thakur2021beir,macavaney-etal-2020-sledge}) or as an initial stage of training before fine-tuning on a target dataset~\cite{li:arxiv2020-parade}. In this section, we describe our experiments training a leading neural re-ranking model on a simulated set of training data that simulate a stricter annotation setup that would have resulted in more discarded queries.

\begin{figure}
\centering
\includegraphics[scale=0.6]{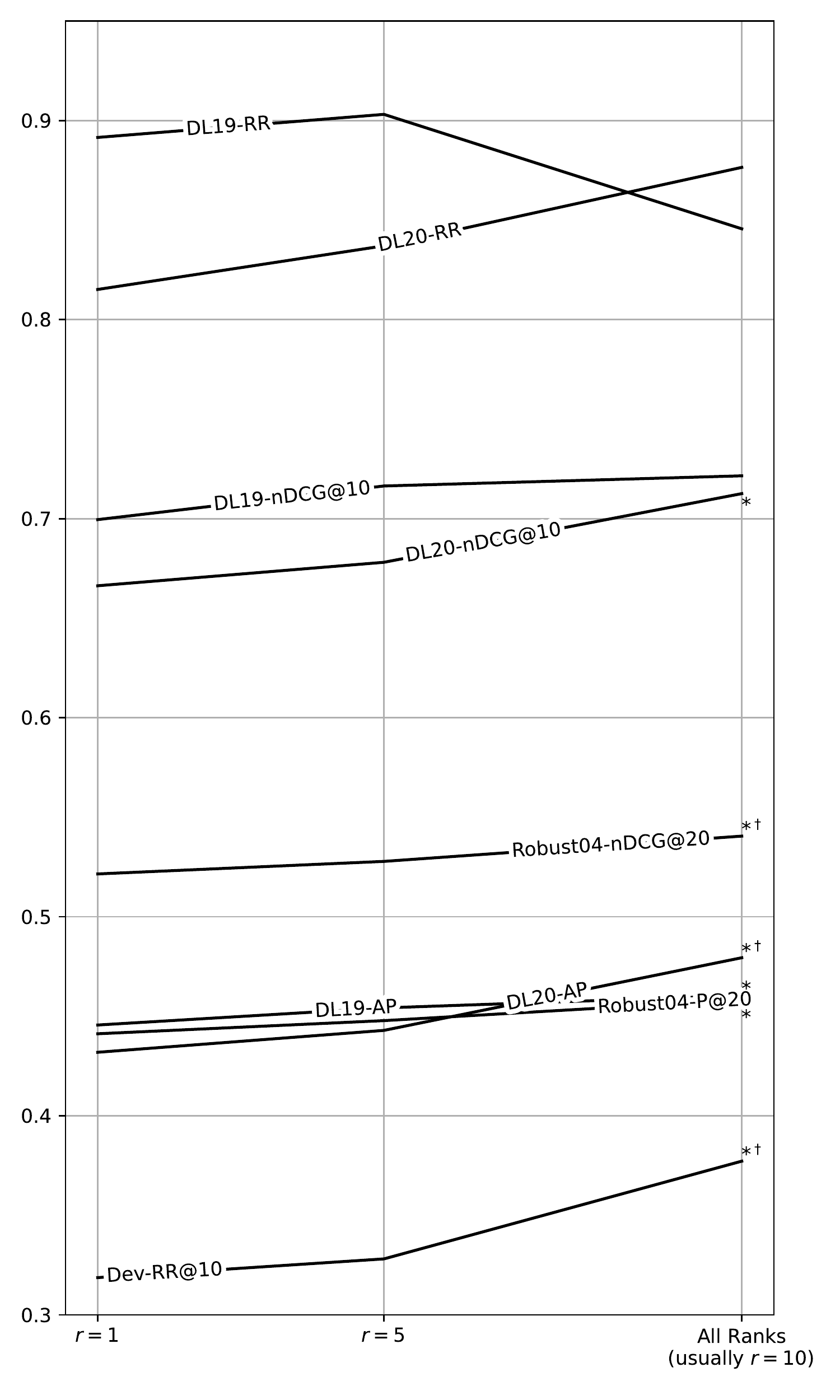}
\vspace{-1.7em}
\caption{Effect of simulated survivorship filtering in monoT5 model training. Significant differences between the simulated settings are indicated with * ($r=1$) and $\dagger$ ($r=5$). Significance is determined using a paired t-test with $p<0.05$ and a Bonferroni correction.}
\label{fig:train}
\end{figure}

\subsection{Experimental Setup}

Akin to our evaluation experiments, we simulate new versions of the MS MARCO dataset where annotators were given fewer documents upon which to find a possible answer. Given a minimum rank~$r$, we filter down the official MS MARCO training triple sequence to only samples where the positive query-document pair appears within the top $r$ documents presented to the annotator.\footnote{We acknowledge that in this setup, the true relevant answers that were observed above $r$ are not considered as \textit{negative} candidates either. Despite this, we prefer our experimental setup because it ensures as fair as possible of a comparison between systems; they are trained on the same data in the same sequence, just potentially with some samples discarded. Further, since the negative documents were sampled from the top 1000 BM25 results for the official training sequence, the rate at which these false negatives would be encountered would be rather low.} Given the high effectiveness and ease of training of the monoT5 model, we select it as a representative ranking model for this experiment. Using both $r=1$ and $r=5$, we train a monoT5 over 256k samples, a mini-batch size of 8, and a learning rate of $5\times10^{-5}$. As the baseline trained on all available data, we use the official model checkpoint released by the monoT5 authors.

We evaluate on four datasets. First, we use MS MARCO Dev (small), a commonly-used subset of 7980 queries from the full MS MARCO dev set. Since it is well-known that this dataset contains incomplete assessments~\cite{arabzadeh2021shallow}, we also test on TREC DL 19~\cite{craswell-2019-overview} and~20~\cite{craswell-2020-overview}, which contain 43 and 54 queries (respectively) from the hidden MS MARCO eval set with deeper relevance judgments (215 and 210 qrels per query on average, respectively, pooled from a variety of systems). Finally, to test the effect on model transferability, we test on TREC Robust04~\cite{voorhees-2004-overview-2}, which contains 249 queries and 1251 qrels per query. We report the official evaluation measures for each dataset. For both MS MARCO-based datasets we use the passage corpus and the official initial ranking set. For Robust04, where there is no standard initial ranking set, we re-rank the top 100 BM25 results, performing MaxP aggregation over sliding passages of the documents.

\subsection{Results}

Figure~\ref{fig:train} presents the results of our training experiments. The figure shows an apparent upward trend in performance as more ranks are considered as candidates for training data. Specifically, we find that across all four datasets, there are significant differences in performance when training with $r=1$ compared to the full dataset. Despite large absolute differences for RR on DL 19 and 20, the measure is statistically unstable~\cite{Ferrante2021TowardsMS} (especially with few queries available for TREC DL), and did not result in significance. In some cases, there is also statistically significant differences between $r=5$ and the full dataset (Dev RR@10, DL20 AP, and Robust04 nDCG@20). Extrapolating this trend suggests that if the MS MARCO assessments gone even deeper than the typical 10 documents (which would result in fewer discarded queries and increased survivability), performance on test collections like TREC DL and TREC Robust may be improved. This observation suggests that there may be value in annotating discarded queries in the MS MARCO \textit{training} set, since it may improve model performance on both MS MARCO and datasets onto which models trained on MS MARCO are transferred.

\section{Discussion and Conclusion}\label{sec:concl}

In this paper, we delved into the problem of survivorship bias in MS MARCO, a popular dataset used for training and evaluating IR systems. We observed that annotators could not identify answers for 38–45\% of queries, which are subsequently discarded in the MS MARCO passage and document ranking tasks. Through an analysis of a sample of the discarded queries, we found that by far the most common reason that queries were discarded was simply that complete answers were presented to the annotators. We then confirmed that by pooling the results of multiple systems, answers to the majority of discarded queries could be identified. This observation clearly indicates that discarded queries are primarily the result of MS MARCO's shallow pooling annotation process. We further found that the discarded queries likely belong to a different distribution than the surviving queries, as evidenced by differences in the query type between the two sets.

Given that a major factor in query survivability is the annotation depth -- had annotators inspected more documents, they likely would have been able to answer more of the queries -- we simulated lower annotation depths to test the effect of the annotation depth on the training and evaluation of systems. We find that there is a marked difference in performance as the annotation depth is reduced. Importantly, we see this across a variety of ranking models, which suggests that the relative order of systems is likely preserved. Nevertheless, this makes the MS MARCO dev set a poor estimator of the \textit{absolute} performance of systems. In terms of training, we found that training on a reduced-depth subset of MS MARCO training samples reduces the performance of a leading neural re-ranking approach -- when testing in-domain, testing with more complete labels, and testing in a zero-shot setting.

We hope that this study serves as a cautionary tale of the potential for survivorship bias when using shallow pooling techniques. Even simply pooling a few candidate documents from a second system could have reduced the number of discarded queries in MS MARCO substantially. Though re-annotating MS MARCO is prohibitively expensive -- especially for the train set -- we recommend that discarded queries be considered in future annotation efforts, such as those at the Deep Learning Tracks or for preference assessment~\cite{arabzadeh2021shallow}. We further note the potential for discarded queries may be useful in pseudo-labeling settings~\cite{lin-etal-2021-batch}, especially considering that we found existing ranking models can often effectively find answers to these queries.

\begin{acks}
We thank Craig Macdonald, Iadh Ounis, Jeff Dalton, Andrew Yates, and anonymous reviewers for helpful feedback on this manuscript. Sean MacAvaney acknowledges EPSRC grant EP/R018634/1: Closed-Loop Data Science for Complex, Computationally- \& Data-Intensive Analytics.
\end{acks}

\bibliographystyle{ACM-Reference-Format}
\bibliography{ir-anthology,acl-anthology,biblio}

\end{document}